\documentstyle[12pt]{article}
\newcounter{eq}
\newcounter{sc}

\def\overleftrightarrow#1{\vbox{\ialign{##\crcr
 $\leftrightarrow$\crcr\noalign{\kern-1pt\nointerlineskip}
 $\hfil\displaystyle{#1}\hfil$\crcr}}}

\newlength{\minitwocolumn}
\begin{document}

%%%%%%%%%%%%%%%%%%%%%%%%%%%%%%%%%%%%%%%%%%%%%%%%%%%%%%%%%%%%%%%%%%
%%%%%%%%%%%%%%%%%%%%%%%% Title %%%%%%%%%%%%%%%%%%%%%%%%%%%%%%%%%%%
%%%%%%%%%%%%%%%%%%%%%%%%%%%%%%%%%%%%%%%%%%%%%%%%%%%%%%%%%%%%%%%%%%
\begin{flushright}
DPUR/TH/38\\
May, 2013\\
%hep-th/070****\\
\end{flushright}
\vspace{20pt}

%\magnification=\magstep1
\pagestyle{empty}
\baselineskip15pt
%\font\cmssB=cmss17
%\font\cmssS=cmss10

\begin{center}
{\large\bf Classically Scale-invariant B-L Model and Conformal Gravity
\vskip 1mm }

\vspace{20mm}
Ichiro Oda \footnote{E-mail address:\ ioda@phys.u-ryukyu.ac.jp
}

\vspace{5mm}
           Department of Physics, Faculty of Science, University of the 
           Ryukyus,\\
           Nishihara, Okinawa 903-0213, Japan.\\

\end{center}

%\maketitle

\vspace{5mm}
\begin{abstract}
We consider a coupling of conformal gravity to the classically scale-invariant B-L extended
standard model which has been recently proposed as a phenomenologically viable model
realizing the Coleman-Weinberg mechanism of breakdown of the electro-weak symmetry.
As in a globally scale-invariant dilaton gravity, it is also shown in a locally scale-invariant 
conformal gravity that without recourse to the Coleman-Weinberg 
mechanism, the B-L gauge symmetry is broken in the process of spontaneous symmetry breakdown 
of the local scale invariance (Weyl invariance) at the tree level and as a result 
the B-L gauge field becomes massive via the Higgs mechanism. As a bonus of conformal gravity, the
massless dilaton field does not appear and the parameters in front of the non-minimal coupling of gravity 
are completely fixed in the present model. This observation clearly shows that the conformal gravity
has a practical application even if the scalar field does not possess any dynamical degree of freedom 
owing to the local scale symmetry. 
\end{abstract}

\newpage
\pagestyle{plain}
\pagenumbering{arabic}
%\setcounter{page}{1}

%%%%%%%%%%%%%%%%%%%%%%%%%%%%%%%%%%%%%%%%%%%%%%%%%%%%%%%%%%%%%%%%%%
%%%%%%%%%%%%%%%%%%%%%%%% Article %%%%%%%%%%%%%%%%%%%%%%%%%%%%%%%%%
%%%%%%%%%%%%%%%%%%%%%%%%%%%%%%%%%%%%%%%%%%%%%%%%%%%%%%%%%%%%%%%%%%

\rm
%%%%%%%%%%%%%%%%%%%%%%%%%%%%%%%%%%%%%%%%%%%%%%%%%%%%%%%%%%%%%%%%%%%%%
%%%%%%%%%%%%%%%%%%%%%%%%%%%%%%   SEC  1    %%%%%%%%%%%%%%%%%%%%%%%%%%
%%%%%%%%%%%%%%%%%%%%%%%%%%%%%%%%%%%%%%%%%%%%%%%%%%%%%%%%%%%%%%%%%%%%%
\section{Introduction}

The origin of the masses of elementary particles and the existence of different mass scales have been
a big mystery thus far. The recent discovery of a Higgs particle at the LHC \cite{ATLAS, CMS} might open 
a new avenue to resolution of this important problems. In particular, the LHC \cite{ATLAS, CMS} has 
certified that the masses of fermions, gauge bosons and the Higgs particle itself are generated in terms of
spontaneous symmetry breakdown of gauge symmetries. In this sense, it is expected that scalar fields 
play a fundamental role in the mass generation in particle physics via the 
spontaneous symmetry breakdown of some local symmetries. Of course, the final resolution of the origin
of the particle masses must be waited until one would succeed in formulating a quantum theory of gravity
since the masses naturally couple to the graviton and act as a source of gravitational interaction as 
easily seen in the Einstein equations.
 
The existence of different mass scales has led to different naturalness issues such as "Why is the
electro-weak (EW) scale much smaller than the Planck scale?" and "Why is the observed cosmological
constant so tiny compared to the size of vacuum energy calculated in quantum field theory?",
which, together with the existence of cold dark matter,  require us to attempt to construct a new 
theory beyond the standard model (SM). As is well known nowadays, as one of naturalness problems,
we have the gauge hierarchy problem, in other words, what stabilizes the EW scale against large radiative
corrections from high energy physics?, which is difficult to explain within the framework of the SM.
The most popular scenario to resolution of this problem is to appeal to supersymmetry \cite{Martin}   
where quadratic divergences to the Higgs self-energy are precisely canceled by loop effects between bosons
and fermions. However, the LHC null results in searches for supersymmetry may seek for an alternative
scenario to the gauge hierarchy problem. 

In this context, it is worthwhile to recall that Bardeen has already proposed the following idea \cite{Bardeen}: 
Since the SM is classically scale invariant in the absence of the (negative) mass term of the Higgs field,
the trace of the energy-momentum tensor takes the form
%**   C-trace %%%%%%%%%%%%%%%%%%%%%%%%%%%%%%%%%%%%%%%%%%%%%%%%%%%%%%%%%
\begin{eqnarray}
T^\mu _{(c) \mu} = 2 m^2 H^\dagger H.
\label{C-trace}
\end{eqnarray}
%%%%%%%%%%%%%%%%%%%%%%%%%%%%%%%%%%%%%%%%%%%%%%%%%%%%%%%%%%%%%%%%%%%
Then, radiative corrections change this classical vaule to 
%**   Q-trace %%%%%%%%%%%%%%%%%%%%%%%%%%%%%%%%%%%%%%%%%%%%%%%%%%%%%%%%%
\begin{eqnarray}
T^\mu _{(q) \mu} = 2 \delta m^2 H^\dagger H + \sum_k \beta_k {\cal{O}}_k,
\label{Q-trace}
\end{eqnarray}
%%%%%%%%%%%%%%%%%%%%%%%%%%%%%%%%%%%%%%%%%%%%%%%%%%%%%%%%%%%%%%%%%%%
where $\beta_k$ are the beta functions for the coupling constants $\lambda_k$. The classical
scale invariance in the limit $m^2 \rightarrow 0$ implies that with the regularization of the cutoff $\Lambda$
the mass correction $\delta m^2$ must be proportional to not $\Lambda^2$ but $m^2$, 
thereby making it possible for the theory to avoid receiving large radiative
corrections from the high-energy cutoff $\Lambda$. Note that the key point behind this alternative scenario 
is that in addition to the classical scale invariance, the energy scale of the new physics is well
separated from the EW scale and there are no new dynamical degrees of freedom at intermediate
scales between the EW scale and the high energy one which is usually taken to be the Planck scale.

In order to realize the Bardeen's idea explicitly, some models beyond the SM \cite{Meissner, Iso} have already been
made where the key ingredient is to make use of the Coleman-Weinberg mechanism \cite{Coleman} for breaking 
the scale invariance by radiative corrections. Then, it is natural to ask ourselves if one can
break the scale invariance spontaneously at the tree level without recourse to the Coleman-Weinberg 
mechanism at the loop level. Indeed, in the previous study \cite{Oda0}, we have constructed such a
model where the non-minimal coupling term of gravity in the framework of the scalar-tensor theory of 
gravitation \cite{Fujii} induces spontaneous symmetry breakdown of the scale symmetry as well as 
the local U(1) B-L symmetry simultaneously. 
 
Now we wish to point out that this model based on (global) scale symmetry \cite{Oda0} raises three interesting questions:
\begin{enumerate} 
\item Can the spontaneous symmetry breakdown occur in a theory with local scale symmetry? \footnote{In this article, 
we use the terminology such that scale transformation means a global scale transformation 
whereas its local gauge transformation is referred to local scale transformation or Weyl transformation.}  
(it is a natural idea to promote a global symmetry to a local gauge symmetry in quantum field theories.) 
\item Can the dilaton, which emerges as a Nambu-Goldstone boson of broken scale invariance, be removed
from the physical spectrum?  (This might be important phenomenologically since the dilaton, which would mediate
a finite-range, non-Newton force, has not been observed experimentally yet.)
\item Can we fix the parameter $\xi$ appearing in front of the non-minimal coupling term of the gravity in case of
the (global) scale symmetry? (This parameter brings us
some ambiguities in comparing the obtained results with experiments.)
\end{enumerate} 

In this article, we would like to answer these three questions by constructing a concrete model with the properties of
the local scale symmetry and its spontaneous symmetry breakdown on the basis of the B-L extended SM. Surprisingly enough,
it turns out that these questions are closely related to each other and our new model produces an affirmative
answer to them. The structure of this article is the following: In Section 2, we present a new model which accomodates
the Weyl symmetry, derive the equations of motion and prove the Weyl invariance.
In Section 3, we discuss spontaneous symmetry breakdown of the Weyl symmetry. As in the case of the 
scale invariance, it is shown that the local U(1) B-L symmetry is also broken spontaneously at the same time.
We conclude in Section 4.

%%%%%%%%%%%%%%%%%%%%%%%%%%%%%%%%%%%%%%%%%%%%%%%%%%%%%%%%%%%%%%%%%%%%%
%%%%%%%%%%%%%%%%%%%%%%%%%%%%%%   SEC  2    %%%%%%%%%%%%%%%%%%%%%%%%%%
%%%%%%%%%%%%%%%%%%%%%%%%%%%%%%%%%%%%%%%%%%%%%%%%%%%%%%%%%%%%%%%%%%%%%
\section{The model}

Let us start with the following Lagrangian density \footnote{We follow 
notation and conventions by Misner et al.'s textbook \cite{MTW}, for instance, the flat Minkowski metric
$\eta_{\mu\nu} = diag(-, +, +, +)$, the Riemann curvature tensor $R^\mu \ _{\nu\alpha\beta} = 
\partial_\alpha \Gamma^\mu_{\nu\beta} - \partial_\beta \Gamma^\mu_{\nu\alpha} + \Gamma^\mu_{\sigma\alpha} 
\Gamma^\sigma_{\nu\beta} - \Gamma^\mu_{\sigma\beta} \Gamma^\sigma_{\nu\alpha}$, 
and the Ricci tensor $R_{\mu\nu} = R^\alpha \ _{\mu\alpha\nu}$.
The reduced Planck mass is defined as $M_p = \sqrt{\frac{c \hbar}{8 \pi G}} = 2.4 \times 10^{18} GeV$.
Through this article, we adopt the reduced Planck units where we set $c = \hbar = M_p = 1$ though we sometimes recover
the Planck mass $M_p$ for the clarification of explanation. In this units, all quantities become dimensionless. 
Finally, note that in the reduced Planck units, the Einstein-Hilbert Lagrangian density takes the form
${\cal L}_{EH} = \frac{1}{2} \sqrt{-g} R$.}:
%**   Lagr 0  %%%%%%%%%%%%%%%%%%%%%%%%%%%%%%%%%%%%%%%%%%%%%%%%%%%%%%%%%
\begin{eqnarray}
{\cal L}_0 = \sqrt{-g} \left\{ \xi_1 \left[ \frac{1}{6} \Phi^\dagger \Phi R + g^{\mu\nu} (D_\mu \Phi)^\dagger (D_\nu \Phi)\right] 
- \xi_2 \left[\frac{1}{6} H^\dagger H R + g^{\mu\nu} (D_\mu H)^\dagger (D_\nu H)\right]  + L_m \right\},
\label{Lagr 0}
\end{eqnarray}
%%%%%%%%%%%%%%%%%%%%%%%%%%%%%%%%%%%%%%%%%%%%%%%%%%%%%%%%%%%%%%%%%%%
where $\xi_1, \xi_2$ are positive and dimensionless constants. 
The matter part $L_m$ is given by \footnote{The sign of kinetic term of  $A^{(2)}_\mu$ is chosen to be the 
opposite to that of $A^{(1)}_\mu$, whose reason will be explained later.}
%**   Matter Lagr %%%%%%%%%%%%%%%%%%%%%%%%%%%%%%%%%%%%%%%%%%%%%%%%%%%%%%%%%
\begin{eqnarray}
L_m = - \frac{1}{4} g^{\mu\nu} g^{\rho\sigma} F^{(1)}_{\mu\rho} F^{(1)}_{\nu\sigma} 
+ \frac{1}{4} g^{\mu\nu} g^{\rho\sigma} F^{(2)}_{\mu\rho} F^{(2)}_{\nu\sigma} 
- V(H, \Phi)  + L_m^\prime.
\label{Matter Lagr}
\end{eqnarray}
%%%%%%%%%%%%%%%%%%%%%%%%%%%%%%%%%%%%%%%%%%%%%%%%%%%%%%%%%%%%%%%%%%%
Here the potential $V(H, \Phi)$ is defined as  \cite{Iso}
%**   V %%%%%%%%%%%%%%%%%%%%%%%%%%%%%%%%%%%%%%%%%%%%%%%%%%%%%%%%%
\begin{eqnarray}
V(H, \Phi) = \lambda_{H \Phi} (H^\dagger H)(\Phi^\dagger \Phi)
+ \lambda_H (H^\dagger H)^2 + \lambda_{\Phi} (\Phi^\dagger \Phi)^2,
\label{V}
\end{eqnarray}
%%%%%%%%%%%%%%%%%%%%%%%%%%%%%%%%%%%%%%%%%%%%%%%%%%%%%%%%%%%%%%%%%%%
and $L_m^\prime$ denotes the remaining Lagrangian part of the SM sector such as the Yukawa couplings and
the B-L sector such as right-handed neutrinos, which will be ignored in this article since it is irrelevant to
our argument.  \footnote{We assume that $L_m^\prime$ is invariant under the local scale transformation.
This is always possible by introducing the gauge fields for the conformal algebra and, if necessary,
eliminating some of them by gauge conditions and constraints within the framework of gauged
conformal symmetry \cite{SUGRA}.}

For a generic field $\phi$, the covariant derivative $D_\mu$ is defined as \footnote{We have chosen
diagonal definitions of the covariant derivatives for simplicity.} 
%**   Cov-der %%%%%%%%%%%%%%%%%%%%%%%%%%%%%%%%%%%%%%%%%%%%%%%%%%%%%%%%%
\begin{eqnarray}
D_\mu \phi &=& \partial_\mu \phi + i \left[ g_1 Q^Y A^{(1)}_\mu + g_{BL} Q^{BL} A^{(2)}_\mu \right] \phi,
\nonumber\\
(D_\mu \phi)^\dagger &=& \partial_\mu \phi^\dagger -  i \left[ g_1 Q^Y A^{(1)}_\mu + g_{BL} Q^{BL} A^{(2)}_\mu 
\right] \phi^\dagger,
\label{Cov-der}
\end{eqnarray}
%%%%%%%%%%%%%%%%%%%%%%%%%%%%%%%%%%%%%%%%%%%%%%%%%%%%%%%%%%%%%%%%%%%
where $Q^Y$ and $Q^{BL}$ respectively denote the hypercharge and B-L charge whose corresponding gauge fields are written 
as $A^{(1)}_\mu$ and $A^{(2)}_\mu$. The charge assignment for the complex singlet scalar $\Phi$ 
and the Higgs doublet $H$ is $Q^Y(\Phi) = 0, Q^{BL}(\Phi) = 2, Q^Y(H) = \frac{1}{2}, Q^{BL}(H) = 0$. Moreover, the field strengths
for the gauge fields are defined in a usual manner as
%**   F %%%%%%%%%%%%%%%%%%%%%%%%%%%%%%%%%%%%%%%%%%%%%%%%%%%%%%%%%
\begin{eqnarray}
F^{(i)}_{\mu\nu} = \partial_\mu A^{(i)}_\nu - \partial_\nu A^{(i)}_\mu,
\label{F}
\end{eqnarray}
%%%%%%%%%%%%%%%%%%%%%%%%%%%%%%%%%%%%%%%%%%%%%%%%%%%%%%%%%%%%%%%%%%%
where $i = 1, 2$. 

Then, it is easy to see that in the starting Lagrangian density  (\ref{Lagr 0}) we can set the constants $\xi_1, \xi_2$ to be the unity 
by the field redefinitions of $\Phi, H$ and redefinitions of the coupling constants $\lambda_{H \Phi}, \lambda_H, \lambda_{\Phi}$.
Thus, as the Lagrangian density of our model, instead of ${\cal L}_0$ in (\ref{Lagr 0}), let us take the following expression 
in this paper \footnote{It is a general feature in conformal gravity that the sign of $\Phi$-kinetic term is 'wrong' or unphysical sign. 
Related to this fact, the sign of $A^{(2)}_\mu$-kinetic term was selected to be unphysical sign.}:
%**   Lagr   %%%%%%%%%%%%%%%%%%%%%%%%%%%%%%%%%%%%%%%%%%%%%%%%%%%%%%%%%
\begin{eqnarray}
{\cal L} = \sqrt{-g} \left[ \frac{1}{6} \Phi^\dagger \Phi R + g^{\mu\nu} (D_\mu \Phi)^\dagger (D_\nu \Phi)
- \frac{1}{6} H^\dagger H R - g^{\mu\nu} (D_\mu H)^\dagger (D_\nu H)  + L_m \right],
\label{Lagr}
\end{eqnarray}
%%%%%%%%%%%%%%%%%%%%%%%%%%%%%%%%%%%%%%%%%%%%%%%%%%%%%%%%%%%%%%%%%%%

Since all coupling constants in ${\cal L}$ are dimensionless and suitable coefficients in front of the non-minimal
terms are chosen, the Lagrangian density ${\cal L}$ is invariant under the local scale transformation defined as 
%**   Conformal transf %%%%%%%%%%%%%%%%%%%%%%%%%%%%%%%%%%%%%%%%%%%%%%%%%%%%%%%%%
\begin{eqnarray}
g_{\mu\nu} &\rightarrow& \tilde g_{\mu\nu} = \Omega^2(x) g_{\mu\nu},  \quad
g^{\mu\nu} \rightarrow \tilde g^{\mu\nu} = \Omega^{-2}(x) g^{\mu\nu}, \quad
 \nonumber\\
\Phi &\rightarrow& \tilde \Phi = \Omega^{-1}(x) \Phi, \quad
H \rightarrow \tilde H = \Omega^{-1}(x) H,  \quad
A^{(i)}_\mu \rightarrow \tilde A^{(i)}_\mu = A^{(i)}_\mu.
\label{Conformal transf}
\end{eqnarray}
%%%%%%%%%%%%%%%%%%%%%%%%%%%%%%%%%%%%%%%%%%%%%%%%%%%%%%%%%%%%%%%%%%%
Actually, it is straightforward to prove the Weyl invariance of ${\cal L}$ 
when we use the formulae $\sqrt{-g} = \Omega^{-4} \sqrt{- \tilde g}$ and
%**   Curvature %%%%%%%%%%%%%%%%%%%%%%%%%%%%%%%%%%%%%%%%%%%%%%%%%%%%%%%%%
\begin{eqnarray}
R = \Omega^2 ( \tilde R + 6 \tilde \Box f - 6 \tilde g^{\mu\nu} \partial_\mu f \partial_\nu f ),
\label{Curvature}
\end{eqnarray}
%%%%%%%%%%%%%%%%%%%%%%%%%%%%%%%%%%%%%%%%%%%%%%%%%%%%%%%%%%%%%%%%%%%
with being  defined as $f = \log \Omega$ and $\tilde \Box f = \frac{1}{\sqrt{- \tilde g}} 
\partial_\mu (\sqrt{- \tilde g} \tilde g^{\mu\nu} \partial_\nu f) = \tilde g^{\mu\nu} 
\tilde \nabla_\mu \tilde \nabla_\nu f$.

Now let us derive equations of motion for later convenience. 
The variation of (\ref{Lagr}) with respect to the metric tensor produces the Einstein equations   
%**   Einstein eq %%%%%%%%%%%%%%%%%%%%%%%%%%%%%%%%%%%%%%%%%%%%%%%%%%%%%%%%%
\begin{eqnarray}
\frac{1}{3} (\Phi^\dagger \Phi - H^\dagger H) G_{\mu\nu} = T_{\mu\nu} + T^{(\Phi)}_{\mu\nu} + T^{(H)}_{\mu\nu}
- \frac{1}{3} ( g_{\mu\nu} \Box - \nabla_\mu \nabla_\nu ) (\Phi^\dagger \Phi - H^\dagger H),
\label{Einstein eq}
\end{eqnarray}
%%%%%%%%%%%%%%%%%%%%%%%%%%%%%%%%%%%%%%%%%%%%%%%%%%%%%%%%%%%%%%%%%%%
where d'Alembert operator $\Box$ is as usual defined as $\Box (\Phi^\dagger \Phi) = \frac{1}{\sqrt{-g}} \partial_\mu
(\sqrt{-g} g^{\mu\nu} \partial_\nu (\Phi^\dagger \Phi)) = g^{\mu\nu} \nabla_\mu \nabla_\nu (\Phi^\dagger \Phi)$
and the Einstein tensor is $G_{\mu\nu} = R_{\mu\nu} - \frac{1}{2} g_{\mu\nu} R$.
Here the energy-momentum tensors $T_{\mu\nu}, \ T^{(\Phi)}_{\mu\nu}, T^{(H)}_{\mu\nu}$ are respectively defined as
%**   Energy-momentum %%%%%%%%%%%%%%%%%%%%%%%%%%%%%%%%%%%%%%%%%%%%%%%%%%%%%%%%%
\begin{eqnarray}
T_{\mu\nu} &=& - \frac{2}{\sqrt{-g}} \frac{\delta(\sqrt{-g} L_m)}{\delta g^{\mu\nu}}  \nonumber\\
&=&  - \sum_{i=1}^2 (-1)^i \left( F^{(i)}_{\mu\rho} F^{(i) \rho}_\nu - \frac{1}{4} g_{\mu\nu} F^{(i)}_{\rho\sigma} F^{(i)\rho\sigma} \right)
- g_{\mu\nu} V(H, \Phi),  \nonumber\\
T^{(\Phi)}_{\mu\nu} &=& - \frac{2}{\sqrt{-g}} \frac{\delta}{\delta g^{\mu\nu}} [ \sqrt{-g} g^{\rho\sigma} (D_\rho \Phi)^\dagger 
(D_\sigma \Phi) ]  \nonumber\\
&=&  - 2 (D_{(\mu} \Phi)^\dagger (D_{\nu)} \Phi) + g_{\mu\nu} (D_{\rho} \Phi)^\dagger (D^{\rho} \Phi), \nonumber\\
T^{(H)}_{\mu\nu} &=& - \frac{2}{\sqrt{-g}} \frac{\delta}{\delta g^{\mu\nu}} [ - \sqrt{-g} g^{\rho\sigma} (D_\rho H)^\dagger 
(D_\sigma H) ]  \nonumber\\
&=&  2 (D_{(\mu} H)^\dagger (D_{\nu)} H) - g_{\mu\nu} (D_{\rho} H)^\dagger (D^{\rho} H),
\label{Energy-momentum}
\end{eqnarray}
%%%%%%%%%%%%%%%%%%%%%%%%%%%%%%%%%%%%%%%%%%%%%%%%%%%%%%%%%%%%%%%%%%%
where we have used notation of symmetrization $A_{(\mu} B_{\mu)} = \frac{1}{2} (A_\mu B_\nu + A_\nu B_\mu)$. 

Next, taking the variation with respect to $\Phi^\dagger$ leads to the following equation:
%**   Phi eq %%%%%%%%%%%%%%%%%%%%%%%%%%%%%%%%%%%%%%%%%%%%%%%%%%%%%%%%%
\begin{eqnarray}
\frac{1}{6} \Phi R  - \frac{1}{\sqrt{-g}} D_\mu (\sqrt{-g} g^{\mu\nu} D_\nu \Phi) - \lambda_{H \Phi} (H^\dagger H) \Phi
- 2 \lambda_\Phi (\Phi^\dagger \Phi) \Phi = 0.
\label{Phi eq}
\end{eqnarray}
%%%%%%%%%%%%%%%%%%%%%%%%%%%%%%%%%%%%%%%%%%%%%%%%%%%%%%%%%%%%%%%%%%%
Similarly, the equation of motion for $H^\dagger$ reads
%**   H eq %%%%%%%%%%%%%%%%%%%%%%%%%%%%%%%%%%%%%%%%%%%%%%%%%%%%%%%%%
\begin{eqnarray}
- \frac{1}{6} H R + \frac{1}{\sqrt{-g}} D_\mu (\sqrt{-g} g^{\mu\nu} D_\nu H) - \lambda_{H \Phi} (\Phi^\dagger \Phi) H
- 2 \lambda_H (H^\dagger H) H = 0.
\label{H eq}
\end{eqnarray}
%%%%%%%%%%%%%%%%%%%%%%%%%%%%%%%%%%%%%%%%%%%%%%%%%%%%%%%%%%%%%%%%%%%
Finally, taking the variation with respect to the gauge fields $A^{(i)}_\mu$ produces the "Maxwell" equations
%**   Maxwell eq %%%%%%%%%%%%%%%%%%%%%%%%%%%%%%%%%%%%%%%%%%%%%%%%%%%%%%%%%
\begin{eqnarray}
\nabla_\rho F^{(1)\mu\rho} &=& \frac{1}{2} i g_1 \left[ H^\dagger (D^\mu H) - H (D^\mu H)^\dagger \right], \nonumber\\
\nabla_\rho F^{(2)\mu\rho} &=& 2 i g_{BL} \left[ \Phi^\dagger (D^\mu \Phi) - \Phi (D^\mu \Phi)^\dagger \right].
\label{Maxwell eq}
\end{eqnarray}
%%%%%%%%%%%%%%%%%%%%%%%%%%%%%%%%%%%%%%%%%%%%%%%%%%%%%%%%%%%%%%%%%%%

Now we are ready to prove that with the help of the equation  (\ref{H eq}) for $H$, the equation of motion  (\ref{Phi eq})
for $\Phi$ follows from 
the trace of the Einstein equations, which means that the equation of motion  (\ref{Phi eq}) for $\Phi$ field contain no new dynamical 
information owing to the local scale symmetry.
To do that, let us first take the trace of the Einstein equations (\ref{Einstein eq}) which reads
%**   Trace-Einstein eq %%%%%%%%%%%%%%%%%%%%%%%%%%%%%%%%%%%%%%%%%%%%%%%%%%%%%%%%%
\begin{eqnarray}
\frac{1}{3} (\Phi^\dagger \Phi - H^\dagger H) R = 4 V - 2  (D_\mu \Phi)^\dagger (D^\mu \Phi) 
+ 2 (D_\mu H)^\dagger (D^\mu H) + \Box (\Phi^\dagger \Phi - H^\dagger H).
\label{Trace-Einstein eq}
\end{eqnarray}
%%%%%%%%%%%%%%%%%%%%%%%%%%%%%%%%%%%%%%%%%%%%%%%%%%%%%%%%%%%%%%%%%%%
Next, using the relation
%**   Relation %%%%%%%%%%%%%%%%%%%%%%%%%%%%%%%%%%%%%%%%%%%%%%%%%%%%%%%%%
\begin{eqnarray}
\Box (\Phi^\dagger \Phi) = g^{\mu\nu} [ (D_\mu D_\nu \Phi)^\dagger \Phi 
+ 2  (D_\mu \Phi)^\dagger D_\nu \Phi + \Phi^\dagger D_\mu D_\nu \Phi], 
\label{Relation}
\end{eqnarray}
%%%%%%%%%%%%%%%%%%%%%%%%%%%%%%%%%%%%%%%%%%%%%%%%%%%%%%%%%%%%%%%%%%%
the similar relation for $H$, and the equation of motion (\ref{H eq}) for $H$, it turns out that Eq.  (\ref{Trace-Einstein eq})
can be rewritten as 
%**   Trace-Einstein eq 2 %%%%%%%%%%%%%%%%%%%%%%%%%%%%%%%%%%%%%%%%%%%%%%%%%%%%%%%%%
\begin{eqnarray}
&{}& \Phi^\dagger [\frac{1}{6} \Phi R - g^{\mu\nu} D_\mu D_\nu \Phi - \lambda_{H \Phi} (H^\dagger H) \Phi
-2 \lambda_{\Phi} (\Phi^\dagger \Phi) \Phi]  \nonumber\\
&+&  \Phi [\frac{1}{6} \Phi^\dagger R - g^{\mu\nu} (D_\mu D_\nu \Phi)^\dagger - \lambda_{H \Phi} (H^\dagger H) \Phi^\dagger
-2 \lambda_{\Phi} (\Phi^\dagger \Phi) \Phi^\dagger]  = 0.
\label{Trace-Einstein eq 2}
\end{eqnarray}
%%%%%%%%%%%%%%%%%%%%%%%%%%%%%%%%%%%%%%%%%%%%%%%%%%%%%%%%%%%%%%%%%%%
The terms proportional to $\Phi^\dagger$ are just the Hermitian conjugate of the terms proportional to $\Phi$, 
so each term must vanish separately, thereby proving that the equation of motion (\ref{Phi eq}) for $\Phi$ field
is obtained from the trace of the Einstein equations and the equation of motion for $H$ field.

%%%%%%%%%%%%%%%%%%%%%%%%%%%%%%%%%%%%%%%%%%%%%%%%%%%%%%%%%%%%%%%%%%%%%
%%%%%%%%%%%%%%%%%%%%%%%%%%%%%%   SEC  3    %%%%%%%%%%%%%%%%%%%%%%%%%%
%%%%%%%%%%%%%%%%%%%%%%%%%%%%%%%%%%%%%%%%%%%%%%%%%%%%%%%%%%%%%%%%%%%%%
\section{Spontaneous symmetry breakdown of Weyl symmetry and U(1) B-L symmetry}

Now we are willing to discuss spontaneous symmetry breakdown of Weyl invariance in our
model. In ordinary examples of spontaneous symmetry breakdown in the framework of
quantum field theories, one is accustomed to dealing with a potential which has the shape of 
the Mexican hat type and therefore induces the symmetry breaking in a natural way, but
the same recipe cannot in general be applied to general relativity because of a lack of such a
potential. \footnote{In the case of massive gravity, a similar situation occurs in breaking
the general coordinate invariance spontaneously \cite{Oda}.} Indeed, such a situation has occured
in the case of our previous model with scale symmetry \cite{Oda0}.

On the other hand, the situation at hand is completely different from that of the previous model. 
In the present model  (\ref{Lagr}), there is the local scale invariance, so anyhow one has to fix the gauge 
for many physical applications as in most gauge theories. Then, it is natural to take the following gauge 
condition for the Weyl transformation \footnote{We could replace the Planck mass $M_p$ with 
a more general mass scale $M$, but as will be seen shortly, to get the Einstein-Hilbert term with 
the proper coefficient, $M$ must be almost equal to the Planck mass  $M_p$,
so here we have simply chosen $M = M_p$.} 
%**   Gauge %%%%%%%%%%%%%%%%%%%%%%%%%%%%%%%%%%%%%%%%%%%%%%%%%%%%%%%%%
\begin{eqnarray}
\Phi(x) = \sqrt{3} M_p e^{i \alpha \theta(x)}, 
\label{Gauge}
\end{eqnarray}
%%%%%%%%%%%%%%%%%%%%%%%%%%%%%%%%%%%%%%%%%%%%%%%%%%%%%%%%%%%%%%%%%%%
where $\alpha$ is a constant and $\theta(x)$ is a scalar field, and we have recovered the Planck constant
for the sake of clarity. Incidentally, the gauge transformation leading to this gauge condition turns out to be
%**   Gauge trans%%%%%%%%%%%%%%%%%%%%%%%%%%%%%%%%%%%%%%%%%%%%%%%%%%%%%%%%%
\begin{eqnarray}
g'_{\mu\nu} = \frac{1}{3} (\Phi^\dagger \Phi) g_{\mu\nu}.
\label{Gauge trans}
\end{eqnarray}
%%%%%%%%%%%%%%%%%%%%%%%%%%%%%%%%%%%%%%%%%%%%%%%%%%%%%%%%%%%%%%%%%%%
The reason why we have selected the gauge condition (\ref{Gauge}) is obvious: With this gauge condition, 
the non-minimal term in  (\ref{Lagr}) becomes the standard Einstein-Hilbert term for general relativity
${\cal L}_{EH} = \frac{1}{2} \sqrt{-g} R$.

It is worthwhile to stress that the gauge condition  (\ref{Gauge}) also breaks the (global) scale symmetry by
introducing the Planck mass with mass dimension into the theory. To put differently, we have started with 
a manifestly scale-invariant theory with only dimensionless coupling constants. But in the process
of the gauge fixing of the local scale symmetry, one cannot refrain from introducing the quantity with mass
dimension, which is the Planck mass $M_p$ in the present context, to match the dimensions 
of the equation and consequently the scale invariance is spontaneously broken. Of course,
the absence of a potential which induces symmetry breaking makes it impossible to
investigate a stability of the selected solution, but the very existence of the solution
including the Planck mass with mass dimension justifies the claim that this phenomenon is
nothing but a sort of spontaneous symmetry breakdown. Note that a similar phenomenon can
also be seen in spontaneous compactification in the Kaluza-Klein theories.

As clarified in our previous work \cite{Oda0}, the spontaneous symmetry breakdown of the scale symmetry
simultaneously induces the spontaneous symmetry breakdown of the U(1) B-L symmetry and consequently
the Higgs mechanism for the U(1) B-L symmetry. Indeed, with the gauge condition  (\ref{Gauge}),
the second term in (\ref{Lagr}) is cast to the form
%**   2nd term %%%%%%%%%%%%%%%%%%%%%%%%%%%%%%%%%%%%%%%%%%%%%%%%%%%%%%%%%
\begin{eqnarray}
\sqrt{-g} g^{\mu\nu} (D_\mu \Phi)^\dagger (D_\nu \Phi) 
= 12 g^2_{BL} M_p^2 \sqrt{-g} g^{\mu\nu} B_\mu^{(2)} B_\nu^{(2)},
\label{2nd term}
\end{eqnarray}
%%%%%%%%%%%%%%%%%%%%%%%%%%%%%%%%%%%%%%%%%%%%%%%%%%%%%%%%%%%%%%%%%%%
where we have chosen $\alpha = 2 g_{BL}$ for convenience, and defined a new massive gauge field $B_\mu^{(2)}$ as
%**   B-field %%%%%%%%%%%%%%%%%%%%%%%%%%%%%%%%%%%%%%%%%%%%%%%%%%%%%%%%%
\begin{eqnarray}
B_\mu^{(2)} = A_\mu^{(2)} + \partial_\mu \theta.
\label{B-field}
\end{eqnarray}
%%%%%%%%%%%%%%%%%%%%%%%%%%%%%%%%%%%%%%%%%%%%%%%%%%%%%%%%%%%%%%%%%%%
Note that in the process of spontaneous symmetry breakdown of the scale invariance 
the Nambu-Goldstone boson $\theta$ is absorbed into the gauge field $A_\mu^{(2)}$ corresponding to
the U(1) B-L symmetry as a longitudinal mode and as a result $B_\mu^{(2)}$ acquires a mass, 
which is nothing but the Higgs mechanism.  In other words,
the B-L symmetry is broken at the same time and the same energy scale that the scale symmetry is spontaneously broken.
At this stage, as promised before, we have to comment on the opposite sign of the kinetic term of the gauge field $A_\mu^{(2)}$.
As mentioned in the previous footnote, the sign of $\Phi$-kinetic term is also unphysical sign, so the sign of 
the kinetic term of the gauge field $A_\mu^{(2)}$ must be unphysical to trigger the Higgs mechanism. As a result,
the squared mass of the gauge field  $B_\mu^{(2)}$ becomes negative in  (\ref{2nd term}), which means that it is tachyonic.
However, it will be seen below that this issue can be avoided in a special value of coupling constants if we have another gauge field
with normal mass. 

Putting all this together, with the gauge condition  (\ref{Gauge}), the Lagrangian density  (\ref{Lagr}) can be
written as 
%**   Lagr 2 %%%%%%%%%%%%%%%%%%%%%%%%%%%%%%%%%%%%%%%%%%%%%%%%%%%%%%%%%
\begin{eqnarray}
{\cal L} = \sqrt{-g} \left[ \frac{1}{2} M_p^2 R +  12 g^2_{BL} M_p^2 g^{\mu\nu} B_\mu^{(2)} B_\nu^{(2)}
- \frac{1}{6} H^\dagger H R - g^{\mu\nu} (D_\mu H)^\dagger (D_\nu H)  + L_m \right],
\label{Lagr 2}
\end{eqnarray}
%%%%%%%%%%%%%%%%%%%%%%%%%%%%%%%%%%%%%%%%%%%%%%%%%%%%%%%%%%%%%%%%%%%
where the matter part $L_m$ now reads
%**   Matter Lagr 2 %%%%%%%%%%%%%%%%%%%%%%%%%%%%%%%%%%%%%%%%%%%%%%%%%%%%%%%%%
\begin{eqnarray}
L_m = - \frac{1}{4} g^{\mu\nu} g^{\rho\sigma} F^{(1)}_{\mu\rho} F^{(1)}_{\nu\sigma} 
+ \frac{1}{4} g^{\mu\nu} g^{\rho\sigma} F^{\prime (2)}_{\mu\rho} F^{\prime (2)}_{\nu\sigma} 
- V(H).
\label{Matter Lagr 2}
\end{eqnarray}
%%%%%%%%%%%%%%%%%%%%%%%%%%%%%%%%%%%%%%%%%%%%%%%%%%%%%%%%%%%%%%%%%%%
Here we have defined the field strength $F^{\prime (2)}_{\mu\nu} = \partial_\mu B^{(2)}_\nu - \partial_\nu B^{(2)}_\mu$ 
and the potential $V(H)$ is given by
%**   V 2 %%%%%%%%%%%%%%%%%%%%%%%%%%%%%%%%%%%%%%%%%%%%%%%%%%%%%%%%%
\begin{eqnarray}
V(H) = 9 M_p^4 \lambda_\Phi + 3 M_p^2 \lambda_{H \Phi} (H^\dagger H)
+ \lambda_H (H^\dagger H)^2.
\label{V 2}
\end{eqnarray}
%%%%%%%%%%%%%%%%%%%%%%%%%%%%%%%%%%%%%%%%%%%%%%%%%%%%%%%%%%%%%%%%%%%

For spontaneous symmetry breakdown of the conventional EW symmetry, let us assume
%**   EW SSB %%%%%%%%%%%%%%%%%%%%%%%%%%%%%%%%%%%%%%%%%%%%%%%%%%%%%%%%%
\begin{eqnarray}
\lambda_{H \Phi} < 0, \quad \lambda_H > 0.
\label{EW SSB}
\end{eqnarray}
%%%%%%%%%%%%%%%%%%%%%%%%%%%%%%%%%%%%%%%%%%%%%%%%%%%%%%%%%%%%%%%%%%%
Then, parametrizing $H^T = (0, v + h) e^{i \varphi}$, up to a cosmological constant 
the potential is reduced to the form
%**   Pot %%%%%%%%%%%%%%%%%%%%%%%%%%%%%%%%%%%%%%%%%%%%%%%%%%%%%%%%%
\begin{eqnarray}
V(H) = \frac{1}{2} m_h^2 h^2 + \sqrt{2 \lambda_H} m_h h^3 + \lambda_H h^4, 
\label{Pot}
\end{eqnarray}
%%%%%%%%%%%%%%%%%%%%%%%%%%%%%%%%%%%%%%%%%%%%%%%%%%%%%%%%%%%%%%%%%%%
where we have defined
%**   Parameters %%%%%%%%%%%%%%%%%%%%%%%%%%%%%%%%%%%%%%%%%%%%%%%%%%%%%%%%%
\begin{eqnarray}
v^2 = \frac{3}{2} \frac{|\lambda_{H \Phi}|}{\lambda_H} M_p^2 = \frac{m_h^2}{8 \lambda_H}, \quad
m_h^2 = 12 |\lambda_{H \Phi}| M_p^2.
\label{Parameters}
\end{eqnarray}
%%%%%%%%%%%%%%%%%%%%%%%%%%%%%%%%%%%%%%%%%%%%%%%%%%%%%%%%%%%%%%%%%%%
As the result of the EW symmetry breaking, the Higgs field becomes massive and
the non-minimal term changes like
%**   Lagr 3 %%%%%%%%%%%%%%%%%%%%%%%%%%%%%%%%%%%%%%%%%%%%%%%%%%%%%%%%%
\begin{eqnarray}
- \frac{1}{6} H^\dagger H R = - \frac{1}{6} (v + h)^2 R = - \frac{1}{6} v^2 R + \cdots,
\label{Lagr 3}
\end{eqnarray}
%%%%%%%%%%%%%%%%%%%%%%%%%%%%%%%%%%%%%%%%%%%%%%%%%%%%%%%%%%%%%%%%%%%
where $\cdots$ denotes interactions between the Higgs particle and the graviton.
Since the scale of the VEV of the EW symmetry breaking, $v$ is much smaller than the Planck
mass, we can safely neglect this term compared to the Einstein-Hilbert term in  (\ref{Lagr 2}).

Next, let us consider the kinetic term of $H$ field.  It is straightforward to rewrite it in the form
%**   H-kinetic term %%%%%%%%%%%%%%%%%%%%%%%%%%%%%%%%%%%%%%%%%%%%%%%%%%%%%%%%%
\begin{eqnarray}
- \sqrt{-g} g^{\mu\nu} (D_\mu H)^\dagger (D_\nu H) 
= - \sqrt{-g} g^{\mu\nu} \left[ \partial_\mu h \partial_\nu h + \frac{g_1^2}{4} (v + h)^2 B_\mu^{(1)} B_\nu^{(1)}
\right],
\label{H-kinetic term}
\end{eqnarray}
%%%%%%%%%%%%%%%%%%%%%%%%%%%%%%%%%%%%%%%%%%%%%%%%%%%%%%%%%%%%%%%%%%%
where we have defined a new massive gauge field $B_\mu^{(1)}$ as
%**   B-field2 %%%%%%%%%%%%%%%%%%%%%%%%%%%%%%%%%%%%%%%%%%%%%%%%%%%%%%%%%
\begin{eqnarray}
B_\mu^{(1)} = A_\mu^{(1)} + \frac{2}{g_1} \partial_\mu \varphi.
\label{B-field2}
\end{eqnarray}
%%%%%%%%%%%%%%%%%%%%%%%%%%%%%%%%%%%%%%%%%%%%%%%%%%%%%%%%%%%%%%%%%%%
Thus, up to a cosmological constant, the Lagrangian density (\ref{Lagr}) can be
cast to  
%**   Lagr 4 %%%%%%%%%%%%%%%%%%%%%%%%%%%%%%%%%%%%%%%%%%%%%%%%%%%%%%%%%
\begin{eqnarray}
{\cal L} &=& \sqrt{-g} \Bigl[ \frac{1}{2} M_p^2 R + \frac{1}{4} (F_{\mu\nu}^{\prime (2)})^2
+ 12 g_{BL}^2 M_p^2 B_\mu^{(2)} B^{(2) \mu} - \frac{1}{6} (v + h)^2 R  
- \frac{1}{4} (F_{\mu\nu}^{\prime (1)})^2 
\nonumber\\
&-& \frac{g_1^2}{4}  (v + h)^2 B_\mu^{(1)} B^{(1) \mu}  - \partial_\mu h \partial^\mu h 
- \frac{1}{2} m_h^2 h^2 - \sqrt{2 \lambda_H} m_h h^3 - \lambda_H h^4 \Bigr],
\label{Lagr 4}
\end{eqnarray}
%%%%%%%%%%%%%%%%%%%%%%%%%%%%%%%%%%%%%%%%%%%%%%%%%%%%%%%%%%%%%%%%%%%
where we have defined the field strength $F^{\prime (i)}_{\mu\nu} = \partial_\mu B^{(i)}_\nu 
- \partial_\nu B^{(i)}_\mu$.

Now we would like to consider the issue of mass spectrum of the massive gauge fields. For this purpose,
it is enough to focus on only the quadratic terms of the gauge fields.
%**   Q-term %%%%%%%%%%%%%%%%%%%%%%%%%%%%%%%%%%%%%%%%%%%%%%%%%%%%%%%%%
\begin{eqnarray}
\frac{1}{\sqrt{-g}} {\cal L}_B = \frac{1}{4} (F_{\mu\nu}^{\prime (2)})^2 + 12 g_{BL}^2 M_p^2 B_\mu^{(2)} B^{(2) \mu} 
- \frac{1}{4} (F_{\mu\nu}^{\prime (1)})^2 - \left(\frac{g_1 v}{2} \right)^2  B_\mu^{(1)} B^{(1) \mu}.  
\label{Q-term}
\end{eqnarray}
%%%%%%%%%%%%%%%%%%%%%%%%%%%%%%%%%%%%%%%%%%%%%%%%%%%%%%%%%%%%%%%%%%%
After some calculations, it turns out that when the coupling constants $g_{BL}$ and $g_1$ satisfy the specific relation
%**   CC relation %%%%%%%%%%%%%%%%%%%%%%%%%%%%%%%%%%%%%%%%%%%%%%%%%%%%%%%%%
\begin{eqnarray}
g_{BL} = \frac{v}{4 \sqrt{3} M_p} g_1,  
\label{CC relation}
\end{eqnarray}
%%%%%%%%%%%%%%%%%%%%%%%%%%%%%%%%%%%%%%%%%%%%%%%%%%%%%%%%%%%%%%%%%%%
the quadratic Lagrangian (\ref{Q-term}) can be written as 
%**   Q-term2 %%%%%%%%%%%%%%%%%%%%%%%%%%%%%%%%%%%%%%%%%%%%%%%%%%%%%%%%%
\begin{eqnarray}
\frac{1}{\sqrt{-g}} {\cal L}_B = - \frac{1}{4} \left[ (F_{\mu\nu}^{(B)})^2 + (F_{\mu\nu}^{(B) *})^2 \right]
- \left(\frac{g_1 v}{2} \right)^2  \left( B_\mu^2 + B_\mu^{* 2} \right), 
\label{Q-term2}
\end{eqnarray}
%%%%%%%%%%%%%%%%%%%%%%%%%%%%%%%%%%%%%%%%%%%%%%%%%%%%%%%%%%%%%%%%%%%
where  we have defined 
%**   Def of gauge fields %%%%%%%%%%%%%%%%%%%%%%%%%%%%%%%%%%%%%%%%%%%%%%%%%%%%%%%%%
\begin{eqnarray}
B_\mu &=& \frac{B_\mu^{(1)} + i B_\mu^{(2)}}{\sqrt{2}}, \quad
B_\mu^* = \frac{B_\mu^{(1)} - i B_\mu^{(2)}}{\sqrt{2}},  \nonumber\\
F_{\mu\nu}^{(B)} &=& \partial_\mu B_\nu - \partial_\nu B_\mu, \quad
F_{\mu\nu}^{(B) *} = \partial_\mu B_\nu^* - \partial_\nu B_\mu^*.
\label{Def of gauge fields}
\end{eqnarray}
%%%%%%%%%%%%%%%%%%%%%%%%%%%%%%%%%%%%%%%%%%%%%%%%%%%%%%%%%%%%%%%%%%%
This Lagrangian density implies that the non-Hermitian gauge fields $B_\mu, B_\mu^*$ are massive
fields with the mass $M_B = 2 \sqrt{3} g_{BL} M_p$. In this way, one can evade the problem of a tachyonic mass
if one allows the existence of the gauge fields $B_\mu, B_\mu^*$ in the mass spectrum.

Now we are moving to phenomenology of our theory.
By the order estimate of magnitude, $m_h \approx v \approx 10^{-16} M_p$, which requires us 
to take two conditions
%**   Condition %%%%%%%%%%%%%%%%%%%%%%%%%%%%%%%%%%%%%%%%%%%%%%%%%%%%%%%%%
\begin{eqnarray}
\lambda_H \approx 1, \quad |\lambda_{H \Phi}| \approx 10^{-33}.
\label{Condition}
\end{eqnarray}
%%%%%%%%%%%%%%%%%%%%%%%%%%%%%%%%%%%%%%%%%%%%%%%%%%%%%%%%%%%%%%%%%%%
The former condition is the conventional condition of the EW symmetry breakdown, meaning that 
the Higgs self-coupling is strong and in the regime of the order 1 at the low energy. 
On the other hand, the latter condition is the original one in the present
theory.  Note that $|\lambda_{H \Phi}| \approx 10^{-33}$ is much smaller than $|\lambda_{H \Phi}|
\approx 10^{-3}$ which was derived by using the Coleman-Weinberg mechanism in Ref. \cite{Iso}. 
In order to keep such a small coupling constant from radiative corrections, it will necessary that
$\Phi$ sector should be almost decoupled from the EW sector. In fact, the relation  (\ref{CC relation})
shows that the coupling constant $g_{BL}$ is much smaller than the EW coupling constant $g_1$ like
$g_{BL} \approx 10^{-16} g_1$.

Moreover, note that the scale of the B-L symmetry breaking is approximately characterized 
by the mass of $B_\mu, B_\mu^*$,  which is $M_B = \frac{g_1 v}{2} = 2 \sqrt{3} g_{BL} M_p$. 
This relation suggests that the symmetry breakings of the B-L symmetry and the local scale 
symmetry happens around the TeV scale. It is of interest that only the classical analysis
 in the model at hand indicates the breaking of the B-L symmetry around the TeV scale. 
On the other hand, in the case of the model based on global scale symmetry \cite{Oda0},
the model does not impose any strict restriction on the breaking scale of the B-L symmetry.
Thus, for instance,  given $g_{BL} \approx 1$ in the previous model \cite{Oda0}, both the B-L symmetry and 
the (global) scale symmetry are broken near the Planck scale. Incidentally, from the viewpoint of superstring theories, 
it might be natural to conjecture that the scale symmetry is explicitly broken around the Planck scale
due to the existence of massive states stemming from the excited modes of string \cite{Kawamura}.

As one disadvantage of the present model, the massive gauge bosons  $B_\mu, B_\mu^*$ are
non-Hermitian, which seems to be against experiments. However, the troublesome gauge field $B_\mu^{(2)}$
is associated with the B-L gauge field $A_\mu^{(2)}$ and the Nambu-Goldstone boson $\theta$ coming from
dilaton sector, and has a completely different origin compared to the normal, massive EM gauge field  $B_\mu^{(1)}$,
so we expect that radiative corrections might give a larger contribution to  $A_\mu^{(2)}$ and make it decouple 
from the mass spectrum in the low energy.

%%%%%%%%%%%%%%%%%%%%%%%%%%%%%%%%%%%%%%%%%%%%%%%%%%%%%%%%%%%%%%%%%%%%%
%%%%%%%%%%%%%%%%%%%%%%%%%%%%%%   SEC  4    %%%%%%%%%%%%%%%%%%%%%%%%%%
%%%%%%%%%%%%%%%%%%%%%%%%%%%%%%%%%%%%%%%%%%%%%%%%%%%%%%%%%%%%%%%%%%%%%
\section{Conclusion}

In this article, we have considered a coupling of conformal gravity to a classically scale-invariant B-L
extension of standard model. Our classical action exhibits a local scale symmetry which is spontaneously broken
and consequently it becomes equivalent to Einstein's general relativity coupled to the extended standard model.
Because of the scale symmetry, mass terms in the action are forbidden so that they must be somehow generated in a dynamical way.
In our formalism, the scale symmetry is spontaneously broken at the same time as the spontaneous symmetry breakdown
of the U(1) B-L symmetry around the TeV scale, and as a result the non-trivial potential is generated, leading to the usual electro-weak 
symmetry breaking, by which the Higgs field and fermions etc. acquire their masses. 

Although we have taken account of a specific model, it is obvious to apply our idea to any model of the standard model extensions 
which accomodates the Weyl invariance. Three key points for realizing our mechanism 
are 1) there is the local scale invariance (or the scale invariance), 2) scalar fields associated with some gauge symmetries 
couple to the gravity in a non-minimal way and 3) the scale invariance is spontaneously
broken.  Thus, it seems that our mechanism can be generalized to non-abelian gauge symmetries as well.
In our approach, we implicitly assume that there is no new physics between the electroweak and Planck scales, and in a sense 
the electroweak scale is determined by Planck physics. Then, it is physically reasonable to incorporate the gravity sector 
into the action.  

One might be tempted to change the order of spontaneous symmetry breakdown when several symmetries coexist in a theory.
In our case, one could suppose that the EW symmetry breaking happens first, and then the scale symmetry is spontaneously broken.
However, it is the scale symmetry breaking that precedes all the symmetry breakings since all quantum field theories must
in principle contain the gravity from the beginning.   

Our consideration in this article is confined to the classical analysis. The equivalence between the conformal gravity and the 
Einstein's general gravity might not survive at the quantum level due to quantum fluctuations and in particular to the FP ghosts
associated with the Weyl symmetry. In the future, we wish to study the quantum aspects of the present formalism. 

In the textbook of the scalar-tensor gravity \cite{Fujii}, at the end of the Appendix H, it is written that "$\cdots$ 
though it is not clear that this (conformal gravity) is useful in any practical application. The scalar
field does not have any dynamical degree of freedom, like a gauge function in the theory of a vector field." 
However, we think that the idea that gravitation arises from spontaneous symmtery breakdown of the
Weyl symmetry in the conformal gravity is very appealing and deserves further investigation.

%%%%%%%%%%%%%%%%%%%%%%%%%%%%%%%%%%%%%%%%%%%%%%%%%%%%%%%%%%%%%%%%%%
%%%%%%%%%%%%%%%%%%%%%%%% Acknowledgements %%%%%%%%%%%%%%%%%%%%%%%%%%%%%
%%%%%%%%%%%%%%%%%%%%%%%%%%%%%%%%%%%%%%%%%%%%%%%%%%%%%%%%%%%%%%%%%%
\begin{flushleft}
{\bf Acknowledgements}
\end{flushleft}
We are grateful to Y. Kawamura for valuable discussions. We thank S. Deser for information 
on a paper of conformal gravity \cite{Deser}.
This work is supported in part by the Grant-in-Aid for Scientific 
Research (C) Nos. 22540287 and 25400262 from the Japan Ministry of Education, Culture, 
Sports, Science and Technology.

%%%%%%%%%%%%%%%%%%%%%%%% reference %%%%%%%%%%%%%%%%%%%%%%%%%%%%%%%
%%%%%%%%%%%%%%%%%%%%%%%%%%%%%%%%%%%%%%%%%%%%%%%%%%%%%%%%%%%%%%%%%%

\end{document}